# A TAXONOMY OF CIRCULAR ECONOMY INDICATORS

Authors' names and affiliations:
Michael Saidani[1], Bernard Yannou[1], Yann Leroy[1], François Cluzel[1], Alissa Kendall[2]
[1]Laboratoire Genie Industriel, CentraleSupélec, Université Paris-Saclay, France
[2]Department of Civil and Environmental Engineering, University of California, Davis, USA
Contact author: michael.saidani@centralesupelec.fr

Abstract:

Implementing circular economy (CE) principles is increasingly recommended as a convenient solution to meet the goals of sustainable development. New tools are required to support practitioners, decision-makers and policy-makers towards more CE practices, as well as to monitor the effects of CE adoption. Worldwide, academics, industrialists and politicians all agree on the need to use CE-related measuring instruments to manage this transition at different systemic levels. In this context, a wide range of circularity indicators (C-indicators) has been developed in recent years. Yet, as there is not one single definition of the CE concept, it is of the utmost importance to know what the available indicators measure in order to use them properly. Indeed, through a systematic literature review – considering both academic and grey literature – 55 sets of C-indicators, developed by scholars, consulting companies and governmental agencies, have been identified, encompassing different purposes, scopes, and potential usages. Inspired by existing taxonomies of eco-design tools and sustainability indicators, and in line with the CE characteristics, a classification of indicators aiming to assess, improve, monitor and communicate on the CE performance is proposed and discussed. In the developed taxonomy including 10 categories, C-indicators are differentiated regarding criteria such as the levels of CE implementation (e.g. micro, meso, macro), the CE loops (maintain, reuse, remanufacture, recycle), the performance (intrinsic, impacts), the perspective of circularity (actual, potential) they are taking into account, or their degree of transversality (generic, sector-specific). In addition, the database inventorying the 55 sets of C-indicators is linked to an Excel-based query tool to facilitate the selection of appropriate indicators according to the specific user's needs and requirements. This study enriches the literature by giving a first need-driven taxonomy of C-indicators, which is experienced on several use cases. It provides a synthesis and clarification to the emerging and must-needed research theme of C-indicators, and sheds some light on remaining key challenges like their effective uptake by industry. Eventually, limitations, improvement areas, as well as implications of the proposed taxonomy are intently addressed to guide future research on C-indicators and CE implementation.

Highlights:

- There is a growing need to monitor the circular economy transition and to measure its effects.
- 55 sets of circularity indicators (C-indicators) are reviewed and classified.
- A need-driven taxonomy is proposed to clarify their purposes and possible usages.
- An associated selection tool is provided to facilitate the identification of suitable C-indicators.
- The uptake of C-indicators by the industry and other promising challenges are discussed.

Abbreviations:

- CE: Circular economy
- C-indicators: Circularity indicators
- EASAC: European Academies Science Advisory Council
- EC: European Commission
- EEA: European Environment Agency
- EMF: Ellen MacArthur Foundation
- OECD: Organisation for Economic Co-operation and Development
- SD: Sustainable development
- SDI: Sustainable development indicators

# 1. INTRODUCTION

## 1.1. A CIRCULAR ECONOMY IN TRANSITION, FOR THE SAKE OF SUSTAINABLE DEVELOPMENT

In 1987, the Brundtland Commission called for the creation of new ways to assess progress toward sustainable development (SD), resulting in the emergence of a wide variety of sustainable development indicators (SDI) advanced by academics, companies, environmental agencies and governmental organizations. (Hardi and Zdan, 1997; Jesinghaus, 2014). Now, the adoption of circular economy (CE) practices appears as a timely, relevant and practical option to meet the goals of SD. In fact, Schroeder et al. (2018) showed that the implementation of CE approaches can be applied as a "toolbox" for achieving a sizeable number of SD targets. Accordingly, the CE paradigm is being extensively explored by institutions as a possible path to increase the sustainability of our economic system (Elia et al. 2017). To some, e.g. Linder et al. (2017), the ultimate goal of a CE is a SD. Sustainability can be regarded as an abstract concept for which many stakeholders find difficult to create targets for, in the way it can have diverse meanings to different stakeholders (Earley, 2017). Similarly, the analysis of 114 CE-related definitions by Kirchherr et al. (2017) provides a quantitative evidence that CE means also different things to different people. Nonetheless, both concepts need appropriate means of evaluation to forge ahead. Bocken et al. (2017) outline the importance of indicators in taking the circularity to the next level. In fact, advancing the discussion of the CE to a higher level requires to reach a shared understanding and common language (Blomsma and Brennan 2017). For instance, assessment methods such as the use of indicators can play a key role in generating a deeper understanding and integration of the CE, e.g. in helping industrial practitioners setting suitable circular economy targets (i.e. intended and quantified objectives linked to CE-related strategies).

## 1.2. A GROWING NEED FOR CIRCULARITY INDICATORS: HISTORY AND CURRENT ISSUES

The measurement of circularity is at the center of many questions recently raised by researchers, such as: how to measure the progress of the transition towards a CE? (Potting et al. 2016); how should we measure its performance since its objectives – e.g. reduce, reuse, recycle – are substantially different from those in the traditional linear economy? (EASAC, 2016); how is circularity measured in businesses and economies? (Bocken et al. 2017); how should product-level circularity be measured? (Linder et al., 2017). According to the EASAC (2015), companies may lack the information, confidence and capacity to move to CE solutions due to a lack of (i) indicators and targets, (ii) awareness on alternative circular options and economic benefits, and (iii) the existence of skills gaps in the workforce and lack of CE programmes at all levels of education (e.g. in design, engineering, business schools). In fact, information exchange is actually cited as a constraint to the success of CE practices (Winans et al. 2017). Consistently, without an evaluation framework or support from the industry, CE initiatives are not sustained. By conducting an analysis of indicators that may be appropriate for monitoring progress towards a circular economy, the EEA (2016) noticed the current knowledge base on the CE is rather fragmented. The EEA stated that more structured information is thus needed to inform decision-making and to improve circular business investment decisions. This statement concurs with Haas et al. (2015) for who it is imperative to determine the current state of circularity so that one can have a benchmark against which to track improvements.

On this basis, it is now commonly acknowledged that to promote CE, the introduction of monitoring and evaluation tools like indicators to measure and quantify this progress becomes essential (Walker et al. 2018; Acampora et al. 2017; Cayzer et al. 2017; Akerman, 2016; Di Maio and Rem, 2015; Su et al. 2013; Geng et al., 2012). The European Commission has also recognized this need for circularity indicators through its action plan for the CE (EC, 2015a) stating that "to assess progress towards a more circular economy and the effectiveness of action at EU and national level, it is important to have a set of reliable indicators". Additionally, to Wisse (2016), it is important to measure the effectiveness of circular strategies deployed at national, regional, and local levels. As a consequence, more and more attempts at developing indicators for the CE concept are found in the literature (Akerman, 2016). Actually, numerous circularity indicators – as listed in Appendix A – have been developed in the last few years, but in an inconsistent manner regarding their scopes, purposes, and possible applications. Yet, the lack of academic and scientific knowledge on CE indicators is a barrier for further implementation (Akerman, 2016). In this line, Linder et al. (2017) underline an urgent need to carefully review the available solutions for measuring circularity, so as to find solutions to their varying weaknesses, or to identify some complementarities. As a response to this recent growing number of fuzzy and multifaceted C-indicators, a clarification on these indicators would be appreciated to facilitate therefore their dissemination and proper usages.

### 1.3. RESEARCH GAPS AND CONTRIBUTIONS

Dealing with the humongous number of available SDI, Bell and Morse (2008) allege that "now we have developed so many indicators that we are having to ask ourselves, what exactly are we measuring". Without entering into a philosophical debate raised by these authors, in regard to the truth behind the indicators – "your truth is not necessarily my truth, truth is a relative term, and indicators are also relative devices" – which would be way out of the scope here, it makes sense and seems appropriate to clarify here what the existing and so called C-indicators are measuring exactly. Even though the research area on C-indicators is in expansion and is becoming increasingly discussed through the academic literature, there is still a lack of in-depth investigation on their completeness, classification, possible complementary and applicability from an industrial or political perspective. This is partly due to the magnitude of the CE paradigm. Indeed, because of the various and diverse definitions of the CE, some C-indicators are not always very explicit on what they aim to measure, or are not properly positioned e.g. regarding the different principles of the CE. As a consequence, they may be interpreted into many different ways.

The main contribution of this article is therefore to trim the fuzziness on current C-indicators and thus to clear up their utility in an organized, understandable and usable manner. To do so, a proposed taxonomy of C-indicators, adapted to users – either industrialists (e.g. engineers, designers, managers) or policy-makers – and its associated selection tool, are developed and experimented on several use cases published in literature. This actual challenge is in agreement with Behrens et al. (2015) underlining the multitude of existing indicators can create confusion, or by Geisendorf and Pietrulla (2017) advancing the measurement of circularity is considered to play a crucial role in the transition, but there is no prevailing opinion on which operationalization to use. We do understand that finding suitable indicators can be a difficult task in the light of this important number of available C-indicators, but we argue it could be facilitated by the design of an appropriate classification scheme and associated selection tool.

The article is structured in the following way, as illustrated in Figure 1. The specific terms used all along this study are defined hereafter. Section 3 exposes the research methodology to identify, analyse and characterize the C-indicators, as well as to construct this taxonomy. Relevant literature is then discussed in Section 4: the particular interest and applicability of indicators for an enhanced CE are developed, and prior taxonomies in sustainability and eco-design related fields are reviewed. Section 5 details the proposed taxonomy and its associated selection tool tested on several use cases as a first validation of the proposal. Section 6 uses the classification and characterization of C-indicators to discuss and question more in-depth their potentiality in the CE transition, as well as their current limitations. Section 7, hence, opens on future areas of investigation to advance further the CE implementation.

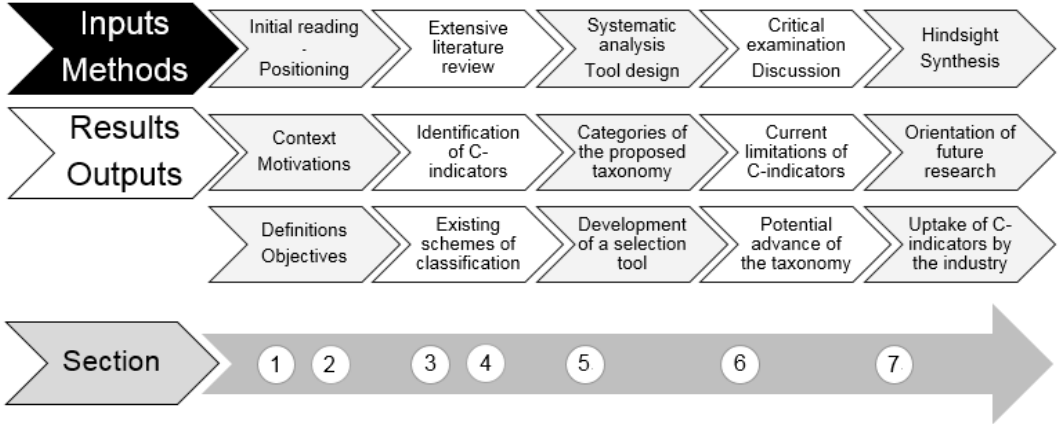

Figure 1 – Synopsis of the article and research process in developing a taxonomy of C-indicators

## 2. DEFINITION OF TERMS – POSITIONING FOR THIS STUDY

The measurement of the circularity performance can lead to several interpretations, as the CE is a fuzzy defined concept. Furthermore, a critical examination of the literature on the CE made by Hass et al. (2015) reveals a lack of precise definitions and criteria for assessing measures to improve the circularity of the economy. Therefore, let us first clarify the terms that are used all along this article.

### 2.1. POSITIONING ON CIRCULAR ECONOMY DEFINITIONS

CE definitions have been comprehensively reviewed by scholars. Sacchi et al. (2018) pointed out the lack of consensus on terminologies and definitions for the CE among scholars, politicians and practitioners investigating the trends, gaps, and convergence of the CE literature – through a sample composed of 327 academic articles. Similarly, Kirchherr et al. (2017) reviewed 114 circular economy definitions which were coded on 17 dimensions. In this article, we refer to the uniting and synthetized definition they proposed: CE is defined as "an economic system that replaces the 'end-of-life' concept with reducing, alternatively reusing, recycling and recovering materials in production/distribution and consumption processes. It operates at the micro level (products, companies, consumers), meso level (eco-industrial parks) and macro level (city, region, nation and beyond), with the aim to accomplish sustainable development, thus simultaneously creating environmental quality, economic prosperity and social equity, to the benefit of current and future generations". Additionally, according to the EMF (2013), the CE is based on three shared principles, which can be summarized as it follows: (i) design out waste and pollution, (ii) keep products and materials in use, and (iii) regenerate natural systems.

### 2.2. INDICATORS AND RELATED SEMANTIC FIELD

A similar story can be told for defining indicators. In fact, the term "indicator" has been defined in various ways in the literature (Park and Kremer, 2017; OECD, 2014; Joung et al. 2013; Singh et al. 2012; EEA, 2003) and there is no one widely agreed upon definition for an indicator. This article adopts the global view of the OECD (2014) where an indicator is defined as "a quantitative or qualitative factor or variable that provides a simple and reliable means to measure achievement, to reflect changes connected to an intervention, or to help assess the performance of a development actor". An indicator framework entails a collection of indicators that "conveys a broader purpose and significance to the individual indicator and provides a comprehensive picture of some entity" (Wisse, 2016). Therefore, indicators simplify information, can help to reveal complex phenomena, and provide an effective tool for measuring progress and performance. Purposes and benefits of the use of indicators are further developed in Section 4.1.

Also, it is important to notice that other terms are found to describe assessment tools, such as "measures", "metrics", 'index", or "indices". In fact, the use of suitable synonyms during the research process (see Section 3.1) is fundamental to ensure a comprehensive identification of existing C-indicators. Even if slight semantic differences are noticed between those terms, most researchers use them interchangeably. As such, for the wording used all along this article, the term indicator is privileged for a better understanding but also because of its generality and common use in the literature. To deal with and manage properly a significant number of indicators, it can be useful to define a classification (i.e. a taxonomy or a typology) of indicators in order to ease their selection process (Lützkendorf and Balouktsi, 2017).

### 2.3. TAXONOMY AND/OR TYPOLOGY

The same goes also for the terms "typology" and "taxonomy" that are often used interchangeably, even if subtle differences can be noticed between these two terms. Typology is the study or system of sorting a large group into smaller groups according to similar features or qualities (Davidson, 1952). Typology creates useful heuristics and provides a systematic basis for comparison. Taxonomy is related to an empirical scheme of classification, suitable for descriptive analysis (Smith, 2002). Although often associated with the biological sciences, taxonomic methods are also employed in numerous disciplines that face the need for categorization schemes. In the scientific literature related to sustainability indicators, eco-design tools or even circular economy business model, the term "taxonomy" is mainly used when it comes to the classification of such indicators, tools or business models, e.g. Rousseaux et al. (2017), Urbinati et al. (2017), Moreno et al. (2016), Bovea and Pérez-Belis (2012). As such, the term 'taxonomy" has been preferred to describe the identification, characterization and classification of C-indicators in the present article.

## 3. MATERIALS AND METHODS

### 3.1. RESEARCH METHODOLOGY

The research method employed in this article is a systematic and extended literature review. The function of a review article is to synthesize literature, to identify research gaps, to highlight emerging patterns, and to recommend new research areas. Here, for the sake of completeness in the identification and screening of C-indicators, the research process includes:

- Combinations of following terms: 'circular economy', 'circularity', 'evaluation', 'assessment' 'measure', 'indicators', 'indices', 'index', and 'metrics' for the database search in title, abstract and keywords fields.
- Academic and non-academic databases: the review was based on both peer-reviewed journals articles or conferences papers and on grey literature. Indeed, in addition to academic literature, complementary sources (e.g. reports, policy communications) were consulted to widely cover the existing knowledge on C-indicators. As such, articles and C-indicators included in this study that are not necessary peer-reviewed – but will be indicated as such, for transparency, in the taxonomy.

Note that this study is limited to C-indicators and related publications in English, and the age of materials reviewed (time coverage) is from the emergence of C-indicators, i.e. 2010 to the submitted date of this research, viz. May 2018. All criteria and associated research items used for the literature review are summarized in Table 1.

Table 1 – Criteria and research filters used to identify C-indicators

| Criteria | | Research item and filter |
|---|---|---|
| Key words | | {circular economy OR circularity} AND {indicators, indices, index, metrics, measure, assessment, evaluation} |
| Databases | Academic | Science Direct, SAGE, Springer, Taylor and Francis, Wiley, Emerald, JSTOR, and Google Scholar. |
| | Non-academic | Web-pages and reports from lobby organizations (e.g. the Ellen MacArthur Foundation), research organisations (e.g. the European Environmental Agency), and governmental agencies (e.g. the European Commission) through Google searches. |
| Language | | English |
| Geographic scope | | Worldwide |
| Publication years (age of material) | | (2000 –) 2010 – May 2018 |

In addition to the systematic literature review carried out to identify the existing C-indicators, a supplementary literature survey was done in parallel – as shown in Figure 2, illustrating the steps of the research process – in order to get inspiration from studies related to the design and proposal of taxonomies previously developed, notably in the fields of sustainability and eco-design. The terms "taxonomy", "typology" and "classification", plus "sustainability" and "eco-design" were hence used as a way of expanding the literature search.

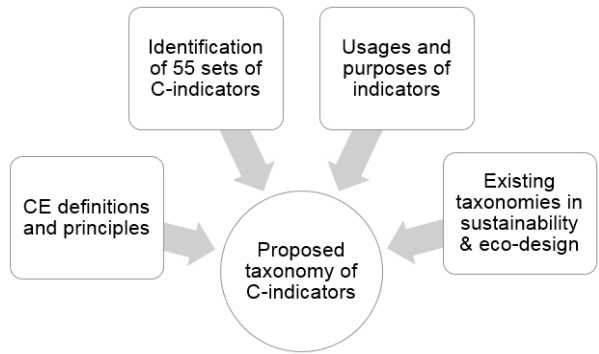

Figure 2 – Sources of inspiration for the proposed taxonomy of C-indicators

### 3.2. MATERIAL INVESTIGATION AND BIBLIOGRAPHIC ANALYSIS

The analysis found 55 sets of C-indicators, coming from 27 journal articles, 2 conference papers, 1 master thesis, 7 technical reports, and 12 websites, tools (n.b. some publications include more than one set of C-indicators). Although the research period starts in 2000, the first specific publication on C-indicators found was from 2010. Since, the increasing number of studies published reveals a clear interest on this topic. Figure 3 shows the distribution of identified sets of C-indicators by origins of development, coverage of CE levels, geographic scope (considering the affiliation of the first author) and time period, confirming the research area of C-indicators is in expansion. Note that among the 20 sets of C-indicators at the micro level of CE, 17 of them have been developed by European contributors. On the contrary, among the 19 sets of C-indicators at the macro level of CE, 9 have been developed by Chinese actors. Indeed, academic publications on the macro level of CE come mostly from China-related cases (Sacchi et al. 2017).

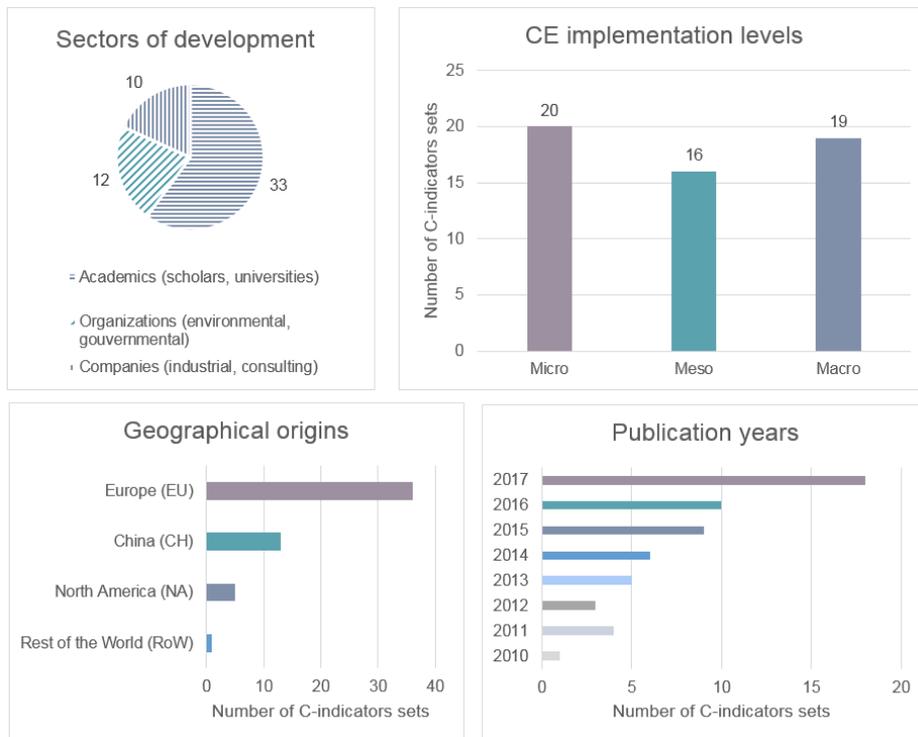

Figure 3 – Bibliographical study: distributions of the C-indicators identified

Then, the information retrieved from exisiting C-indicators – by collecting and analysing all references found, and carefully examining the C-indicators features, principles and possible applications – was structured to propose a classification of C-indicators to facilitate their selection, use and appropriation by industrial practitioners, decision-makers, investors and/or policy-makers interested in moving towards more CE practices.

# 4. LITERATURE BACKGROUND – STATE-OF-THE-ART

## 4.1. INDICATORS: PURPOSES, USAGES, AND BENEFITS

The purposes and advantages offer by the use of indicators have been extensively discussed in the literature. Let us summarize first their principal generic features and benefits, and then especially in regard to the measurement of the CE performance. In fact, indicators have: the ability to summarize, focus and condense the complexity of the dynamic environment to a manageable amount of meaningful knowledge (Singh et al. 2012), that is to say, the potentiality of relaying complex information in a simplified and useful manner (Wisse, 2016); the capability to communicate, raise public awareness on important issues (e.g. potential environmental impacts), and to indicate whether or not targets will be met (EEA, 1999). Indicators can also be used as managerial and policy-making instruments to: report or pilot activities; define goals, quantitative targets, and track progress; arbitrate potential trade-offs and impact transfers; inform investment choices and guide policy-making; communicate externally; support education and training. Last but not least, according to Wass et al. (2014), indicators contribute on the need of short cuts and rules of thumb to support decision-making.

Specifically in regard to the CE, C-indicators can function as a springboard for a transition toward more CE practices (Kalmykova et al. 2018), thanks to their different potential uses (Linder et al. 2017; Arnsperger and Bourg, 2016): as a key performance indicators (to benchmark and compare industries), as product labels (to inform consumer choices), as a basis for regulatory change. For Thomas and Birat (2013), they are essential to capture the stakes of reuse and recycling at the end-of-life of products during decision making. In response to the complexity related to the CE paradigm, considering the interrelations between different actors all along the value chain implied in the CE implementation, C-indicators can provide a standardized language to simplify information exchange, understanding, and thus ease this transition (Verberne, 2016). With such a baseline in place, businesses adopting CE principles can collaborate, advance together, and set targets against which progress towards circularity can be measured. Walker et al. (2018) add that the aim of C-indicators is to inform life cycle design decisions without the need for a full and time-consuming life cycle analysis.

## 4.2. TAXONOMIES IN SUSTAINABILITY AND ECO-DESIGN

Taxonomies facilitate the diffusion of organised knowledge and allow to achieve a higher maturity level on a given concept (Xavier et al. 2015). Different methodologies of classification have been proposed over the last decades, notably in response to the growing number of SDI and eco-design tools. Therefore, before starting the review and classification of C-indicators, let us take inspiration from previous work on developed categorisation schemes of indicators and tools in the fields of sustainable development and eco-design.

### 4.2.1. Classification of sustainable development indicators

Given the number and diversity of sustainability indicators that have been developed, it was becoming more and more difficult for decision- and policy-makers to grab their meaning and relevance. Therefore, some means of structuring and analysing indicators were requested (EEA, 1999) and have emerged. Sustainability indicators often appear classified in regard to three dimensions (Ruiz-Mercado, 2012) e.g. Krajnc and Glavic (2003) who classified 89 indicators according to environmental, economic and social areas; or Sikdar (2003) who created a hierarchical indicator system of these three dimensions depending on how many aspects are measured by the indicator.

Nonetheless, other categorisation schemes have been proposed in the literature. The EEA (2003) classified sustainability-related indicators into five groups: (i) descriptive indicators (including state, pressure or impact variables, expressed in absolute scale); (ii) performance indicators (using the same variables as descriptive indicators but are connected with target values, measuring the distance between the current situation and the desired situation); (iii) efficiency indicators (providing insight into the efficiency of products and processes in terms of – economic and environmental – resources, emissions and waste per unit output); (iv) policy effectiveness indicators (related the actual change of environmental variables to policy efforts); and (v) total welfare indicators. Singh et al. (2012) gave an overview of various SDI and grouped them into the following categories: innovation, knowledge and technology indices; development indices; market and economy based indices; eco-system based

Indices; composite sustainability performance indices for industries; product based sustainability index; sustainability indices for cities; environmental indices for policies, nations and regions; environment indices for industries; social and quality of life based indices; energy based indices; ratings. Additionally, the classification and evaluation of SDI can be done based on the following dimensions: aspects of the sustainability to be measured by indicators; techniques used for development of index like relative or absolute, quantitative or qualitative, unidimensional or multidimensional; measurement of sustainability in terms of input (i.e. means) or output (i.e. ends); clarity and simplicity in its content, purpose and method; availability of data (Singh et al. 2012).

In the meantime, new sets of sustainability indicators have been developed and this classification debate still prevails today (Park and Kremer, 2017). Indeed, despite the variety of available environmental sustainability indicators, Park and Kremer (2017) notice the absence of a commonly accepted categorization framework often creates confusion and inhibits indicator deployment in practice. As a solution, using text-mining techniques, 55 environmental sustainability indicators were extracted from extant literature and grouped into 5 relevant categories to clarify their usage and facilitate their application in companies: (i) environmental impact and chemical release; (ii) pollution from emissions and wastes; (iii) end of life management and chemicals usage related indicators; (iv) raw materials and facility management related indicators; and (v) energy and water management.

#### 4.2.2. Classification of eco-design methods and tools

Following the emergence of eco-design tools that started in the 1990s, several authors have then proposed various classification systems of such tools since 2000. For example, Janin (2000) determined two main categories: environmental assessment and improvement. Hernandez-Pardo et al. (2011) proposed a use-oriented classification regarding three properties: complexity, type, and main function of the eco-design tools.

Bovea and Pérez-Belis (2012) reviewed and classified eco-design assessment tools to facilitate their integration into the product design process. With the intention of providing designers with a guide to selecting the eco-design tool that best fits a specific application, a taxonomy was made according to criteria such as: (i) the method used for the environmental assessment; (ii) the product requirements that need to be integrated in addition to the environmental ones; (iii) whether the tool has a life cycle perspective; (iv) the qualitative and quantitative nature of the environmental evaluation; (v) the stages of the conceptual design process where the tool can be applied; and (vi) whether the tool has been applied to a case study.

According to Rousseaux et al. (2017), all these classifications are generally intended for engineers and designers to help them in their search for ecodesign solutions, but are hardly linkable to the various functions of a company. On this basis, Rousseaux et al. (2017) updated and consolidated the literature review and analysis on eco-design tools by characterizing 629 eco-design tools into a taxonomy, classifying these tools into 22 categories of ecodesign tools and 5 departments in companies. Furthermore, a web-based guide was made available freely to assist companies in finding the most suitable eco-design tools according to their needs.

### 4.3. TAXONOMIES OF CE-RELATED TOOLS

#### 4.3.1. Classification of CE business models and CE design strategies

To inform and help industrials practitioners (e.g. managers, engineers, designers) in selecting or defining their future circular product design and circular business models, researchers have developed taxonomies to identify what business models or design strategies are the most suitable to their needs. Lewandowski (2016) presented an extensive analysis of 20 types of circular business models, identifying and classifying the CE characteristics according to a business model structure, such as the business model canvas. More recently, Urbinati et al. (2017) proposed a taxonomy of CE business models based on the degree of adoption of circularity along two major dimensions: (i) the customer value proposition and interface; and (ii) the value network. Lüdeke-Freund et al. (2018) conducted a review and analysis of 26 existing CE business models, which resulted in a taxonomy, relying on the six main patterns identified for these circular business models: (i) repair and maintenance; (ii) reuse and redistribution; (iii) refurbishment and remanufacturing; (iv) recycling; (v) cascading and repurposing; and (vi) organic feedstock business model patterns.

In a complementary manner, Moreno et al. (2016) proposed a taxonomy of Design for X (DfX) approaches contributing to the implementation of circular design. The taxonomy is based on three DfX approaches: (a) design for resource conservation; (b) design for slowing resource loops; and (c) whole systems design. The taxonomy includes as well five circular design strategies: (i) design for circular supplies; (ii) design for resource conservation; (iii) design for long life use of products; (iv) design for multiple cycles; and (v) design for systems change. Then, a circular design tool (Moreno et al. 2017) was built to present this taxonomy in a non-scientific language with the aim to educate and inspire during the concept development phase. Hollander et al. (2017) depicted a new taxonomy of design approaches for product integrity in a CE, contributing to a deeper understanding on the role of product design in a CE. Thus, their proposed taxonomy provides a basis for comparison and communication that can help product designers make design decisions that will facilitate the transition from a linear to a more CE. Last but not least, Bocken et al. (2016) brought together existing circular product design and business model strategies in the same framework. As such, it provides practical guidance to designers and strategic decision makers in businesses for slowing and closing resource loops.

### 4.3.2. First inventories, reviews and critical analysis of C-indicators

Hass et al. (2015) proposed a set of key indicators to track physical resources, where the degree of circularity of the global economy is measured as the share of actually recycled materials in the total of processed materials. ScoreLCA (2015) identified four stakes in the assessment of CE loops – loops evaluation, loops ranking, loops implementation, loops monitoring – each one with its own methodological stakes. The first objective is to evaluate the environmental, economic and social impacts of loops. The second one is to compare and prioritize different types of loops and to identify the most pertinent solution. The third one is to help the implementation of the selected solutions. Finally, the last objective is to evaluate the evolution of the systems and the performance of the implemented loops. Three major categories of loop assessment methodologies were identified: (i) material flow analysis; (ii) life cycle assessment; and (iii) evaluation and monitoring indicators. Similarly, Wisse (2016) identified three prominent types of frameworks for measuring the CE: (i) material flow accounts; (ii) eco-efficiency indicators; and (iii) hybrid indicators. Reviewing both sustainability and C-indicators, Akerman (2016) established differences between CE core indicators and adapted sustainability indicators. He divided these indicators into five categories: (i) resource productivity; (ii) environmental aspects; (iii) economic opportunities; (iv) social aspects; and (v) waste management. The EASAC (2016) underlined that many available indicators may be appropriate for monitoring progress towards a CE and grouped them into sustainable development, environment, material flow analysis, societal behavior, organizational behavior and economic performance. Yet, only macro-level indicators were considered and other aspects, such as product circularity performance, were not directly considered in these indicators. Banaité and Tamošiūnienė (2016) analysed and provided insights on what should be taking into account when setting up circular economy indicators, through a C-indicators selection model, but at a macro level too.

Before proposing a new C-indicator at a micro level – the PCM (n.b. all acronyms of C-indicators are detailed in Appendix A) – Linder et al. (2017) reviewed five existing product-level C-indicators according to the following criteria, chosen for scientific robustness: construct validity, reliability, transparency, generality, and aggregation principles. Three existing C-indicators – the MCI, CET, and CEIP – to measure product circularity performance have been as well tested by Saidani et al. (2017a) on an industrial case study and then criticized regarding both their practical applicability in industry and compliance with CE principles. Walker et al. (2018) have tested and compared the results given by these three C-indicators with an LCA-based method for the assessment of material circularity. Elia et al. (2017) proposed a taxonomy of methodologies which can be used to measure the environmental effectiveness of CE strategies, based on two factors: (i) the index-based method typology - distinguishing single synthetic indicators and sets of multiple indicators usually divided into several categories; (ii) the parameters to be measured – such as material and energy flow, land use and consumption, and other life cycle based. Pauliuk (2018) proposed a dashboard of C-indicators at the organizational level, completing as such the BS 8001:2017 – standard for implementing CE in organizations – which has weak links to existing accounting and quantitative assessment frameworks, stipulating also that organizations are solely responsible for choosing appropriate CE indicators. The dashboard was set up to select core indicators for the quantitative assessment of CE strategies for organizations and product systems. For instance, for the goal "maintain financial value", the CEI is recommended as a possible indicator, and for the goal "maintain nonfinancial value", the MCI is indicated.

In summary, a complete overview of C-indicators reviewed in the literature is available in Table 2. In total, 28 different C-indicators and associated framework have been reviewed by several authors. In this study, through an extensive literature review, 55 sets of C-indicators have been identified, resulting – to the best of our knowledge – in the most comprehensive analysis of C-indictors so far. They are all listed in Appendix A. The uncounted variety among these indicators provides a relevant basis to start their characterization and classification within an appropriate taxonomy of C-indicators.

Table 2 – Existing reviews, experimentations and critical analysis of C-indicators

| References Authors and Year | Type of publication or journal's name | Type of review and analysis | Number and names of C-indicators considered *(acronyms are detailed in appendix A)* |
|---|---|---|---|
| CIRAIG, 2015 | Environmental Report | Description | 2: MCI, CA |
| Otero, 2015 | Master's Thesis | Description and comparative analysis | 4: MCI, ICT, CECAC, CA |
| Akerman, 2016 | Master's Thesis | Description and comparative analysis | 4: MCI, CA, NCEIS, IPCEIS |
| Wisse, 2016 | Master's Thesis | Description and comparative analysis | 4: FCIM, NCEIS, IPCEIS, EPICE |
| Banaité, 2016 | Journal of security and sustainability issues | Description | 5: BCI, ECEDC, ERCE, DEA, IEDCE |
| Cayzer et al. 2017 | International Journal of Sustainable Engineering | Description and critical analysis, plus experimentation on the developed indicator | 7: CEIP, CET, MCI, EVR, RDI, NCEIS, IPCEIS |
| Saidani et al. 2017a Saidani et al. 2017b | MDPI Recycling Int. Conference Paper | Description, experimentation and critical analysis | 4: MCI, CET, CEIP, CPI |
| Linder et al. 2017 | Journal of Industrial Ecology | Description and critical analysis, plus experimentation on the developed indicator | 6: CEI, MCI, C2C, EVR, RP, PCM |
| Acampora et al. 2017 | Int. Conference Paper | Description and relevance to a specific sector | 8: CEPI, RPI, CEIP, CET, CEI, MCI, EISCE, FCIM |
| Elia et al. 2017 | Journal of Cleaner Production | Description and classification | 13: RPI, CEI, MCI, EVR, HLCAM, RP, FCIM, NCEIS, IPCEIS, ZWI, RCEDI, EPICE, EWMFA |
| Azevedo et al. 2017 | MDPI Resources | Description and classification | 13: RPI, CEI, MCI, EVR, HLCAM, RP, FCIM, NCEIS, IPCEIS, ZWI, RCEDI, EPICE, EWMFA |
| Pauliuk, 2018 | Resources, Conservation and Recycling | Description and classification | 12: CEPI, CEIP, PCM, CEI, MCI, C2C, EVR, RDI, EISCE, NCEIS, IPCEIS, ECEDC |
| Walker et al. 2018 | MDPI Sustainability | Description, experimentation and critical analysis | 6: CEIP, CET, CEI, MCI, C2C, VRE |

## 5. TAXONOMY AND SELECTION TOOL OF C-INDICATORS

In complementarity with existing taxonomies of eco-design tools (e.g. Rousseaux et al. 2017; Bovea and Pérez-Belis, 2012), circular economy business models (Urbinati et al. 2017), and to supplement the first reviews of C-indicators (Pauliuk, 2018; Elia et al. 2017), a taxonomy of C-indicators is proposed and detailed hereafter. In fact, on the grounds of the increasing number of C-indicators developed recently – with different scopes, purposes and usages – the objective is to provide clarity on these indicators, so as to guide CE practitioners towards the right set of indicators, regarding their needs and requirements. As such, the review and analysis of over 50 sets of C-indicators developed and used by academics, companies, environmental organisations or even governmental agencies, have led to their classification into a need-based taxonomy driven by the usage of such indicators, including 10 categories to differentiate and specify these C-indicators, inspired by the CE principles and indicators characteristics. For practical use, a computer-based query tool has been designed to help identifying the most relevant indicators regarding the user's needs, among the databank of 55 sets of C-indicators.

## 5.1. DEFINITION OF THE CATEGORIES FOR THE PROPOSED TAXONOMY

All the 10 categories to classify, differentiate and orient the use of proper C-indicators are summarized in Table 3. Categories from #1 to #4 are specific to the CE paradigm. Categories #5 to #6 are related to the particular usages and fields of application of these C-indicators. Categories #7 and #8 are linked to the basic features of indicators. Category #9 is dedicated to the assessment framework associated to each C-indicator, facilitating for instance its computation. Category #10 specifies the background in which each C-indicator has been developed.

Table 3 – Categories for the proposed taxonomy of C-indicators

| **Categories** *(criteria)* | #1 - **Levels** *(micro, meso, macro)* | #2 - **Loops** *(maintain, reuse/reman, recycle)* | #3 - **Performance** *(intrinsic, impacts)* | #4 - **Perspective** *(actual, potential)* | #5 - **Usages** *(e.g. improvement, benchmarking, communication)* |
|---|---|---|---|---|---|
| | #6 - **Transversality** *(generic, sector-specific)* | #7 - **Dimension** *(single, multiple)* | #8 - **Units** *(quantitative, qualitative)* | #9 - **Format** *(e.g. web-based tool, Excel, formulas)* | #10 - **Sources** *(academics, companies, agencies)* |

First, C-indicators can be divided into micro-level (organization, products, and consumers), meso-level (symbiosis association, industrial parks) and macro-level (city, province, region or country) indicators (Kirchherr et al. 2017). Indeed, CE models and implementations are usually performed at three systemic levels (Acampora et al. 2017; Linder et al. 2017; Ghisellini et al. 2016). As such, these different levels of implementation of CE require the development of different indicator frameworks that measure the CE performance at national, regional, and more local levels (Wisse, 2016; Su et al. 2013; Geng et al. 2012). Examples of C-indicators at these three levels are given in Table 4.

Table 4 – Categorisation of C-indicators according to the micro-, meso- and macro- levels of the CE

| Levels | Applications | Example n°1 | Example n°2 | Example n°3 |
|---|---|---|---|---|
| Macro | Cities, Regions, Nations | Evaluation of CE Development in Cities (ECEDC) | Regional CE Development Index (RCEDI) | National CE Indicator System (NCEIS) |
| Meso | Businesses, Industrial Symbiosis | Sustainable Circular Index (SCI) | Circular Economic Value (CEV) | Circle Assessment (CA) |
| Micro | Products, Components, Materials | Circular Economy Indicator Prototype (CEIP) | Product-Level Circularity Metric (PCM) | Material Circularity Indicator (MCI) |

While the CE only means recycling from the viewpoint of certain actors, it encompasses reducing, reusing and recycling activities for others (Kirchherr et al. 2017). As such, existing C-indicators do not systematically consider all the potential CE loops. On this basis, the second category characterizes the feedback loops taken into consideration by these C-indicators, namely, maintain/prolong, reuse/remanufacturing and recycling, according to the technosphere part of the CE butterfly diagram proposed by the Ellen MacArthur Foundation (EMF, 2015). Note that in the present work, the focus is on the design, operation and end-of-life of industrial systems, made of technical materials (not of biological ones). That is why we are only considering the feedback loops of the right side (i.e. the technological side) of the EMF butterfly diagram,

For the third category, a differentiation is drawn on another central element: the circularity performance, considering whether an intrinsic circularity or a consequential circularity i.e. the effects resulted by such circularity. In fact, some C-indicators measure the inherent circularity (e.g. recirculation rates of resources) while others depict the consequences of CE loops (e.g. on sustainability). In line with Potting et al. (2016), monitoring progress towards a circular economy should address the transition process as well as its effects. More precisely, the EEA (2016) put the emphasis on the fact that assessing the circularity performance should consider both the progress of the process (e.g. resource efficiency, evolution of material consumption) and effects of a CE transition (e.g. evolution of energy consumption, added value of products and services, employment levels). Actually, the measurement of success of the implementation of CE loops should capture economic and environmental benefits (Geisendorf and Pietrulla, 2017). Overall, it has been assumed that benefits of CE adoption outweight the drawbacks regarding sustainable impacts, but sometimes it could result to negative impacts (Geissdoerfer et al. 2017). As such, it is relevant to check and make sure the potential circularity of the systems will lead to effective benefits regarding sustainability, or to know under what conditions. Note that while the wide range of existing sustainability indicators, as reviewed in sub-section 4.2.1 (e.g. the ones which can be computed through life cycle analysis), are not specifically tailored to assess the economic, environmental or social performance of CE-related strategies, the C-indicators considered, in the present taxonomy, to evaluate a consequential circularity, are the ones especially designed for assessing the sustainability performance of CE loops.

The fourth category adds a temporal focus on the CE measurement – retrospective or prospective – and makes a distinction between an actual and a potential circularity. According to Potting et al. (2016), it is useful to evaluate CE transitions by measuring progress before (*ex ante*), during (*ex durante*) and after (*ex post*) the transition process: "An *ex ante* evaluation is relevant to explore whether proposed CE transitions actually have potential to bring about the intended CE effects. *Ex durante* evaluation is important to monitor whether a CE transition process follows the planned route, and leads to the desired effects. *Ex post* evaluations should determine whether the effects of the CE transition process are in accordance with the set goals." Similarly, to Kok et al. (2013), indicators can be used both in the post-process evaluation and in the pre-process design.

For the fifth category, a highlight is made on the possible uses of the available C-indicators. These indicators provide all a certain degree of information on the CE by assessing one or several criteria of the four categories aforementioned. Yet, in accordance with the literature review, there are different potential usages of a C-indicator. The influence degree of indicators is discussed by Lützkendorf and Balouktsi (2017), distinguishing action-oriented indicators that help decision-makers in formulating clear targets and strategies, from information-oriented indicators that help decision-makers in understanding the current situation. Note that the classification of C-indicators in this category is subjected to more subjectivity in the way it demands more interpretations which could vary regarding the users of the C-indicator. For instance, one may deviate some indicators from their initial purposes to better meet their needs. That is the reason why the proposed clustering of indicators in this category only informs on the *a priori* suitable usages of C-indicators, among the four following generic options: (i) information purposes, helping to understand the situation (e.g. tracking progress, benchmarking, identifying areas of improvement); (ii) decision-making purposes, helping to take action (managerial activities, strategies formulation, policy choice); (iii) communication (internally on the achievements to the stakeholders, externally to the public); and (iv) learning (education of workforce, awareness among consumers).

In the sixth category, the transversality of C-indicators among sectors, segments, or industries is indicated. By analogy with the classification of eco-design tools by Rousseaux et al. (2017), generic C-indicators are applicable to all sectors, to any type of company, regardless of its size, location, field or activity. Sector-specific ones are focused on particular sector applications and provide more operational responses. For instance, the PCM developed by Linder et al. (2017) has a high degree of generality and can be applied across different product categories, whereas the BCI developed by Verberne (2016) is designed to assess the circularity performance in the building industry.

The seventh category aims to differentiate the dimensionality of C-indicators. C-indicators of low dimensionality – i.e. that translate circularity into a single number – are useful for managerial decision making (Linder et al. 2017), whereas a high dimensionality can provide a higher degree of intelligibility more suitable for experts – e.g. designers or engineers – in the assessment of product circularity performance (Saidani et al. 2017). Knowing the degree of intelligibility of C-indicators is important to select indicators that are specifically understandable (Lützkendorf and Balouktsi, 2017) for the intended users e.g. a manager non-expert in CE or a research specialized in the CE implementation.

The eighth category gives information of the indicators units, in order to distinguish the C-indicators in terms of their measurability, whether they use a quantitative or qualitative approach. The units used to calculate circularity are a fundamental aspect of any C-indicator (Linder et al. 2017). Units among the sets of C-indicators identified in the proposed taxonomy include different types such as: mass, time (duration in use), intensity (emission, energy, and consumption), return on investment (savings, profit), availability (resources use, recycling rates in percentage). In fact, measuring progress of the CE transition means gathering quantitative, semi-quantitative and/or qualitative data and compiling them into indicators which provide meaningful information.

The ninth level examines the format of the assessment framework associated to the C-indicators in order to ease their calculation. It has been found that the C-indicators are linked whether to formulas to compute manually (the most common option) or to computational tool (including dynamic excel spreadsheet, web-based tool, or other softwares).

Finally, because these C-indicators have been developed by various kind of actors – (i) academia; (ii) industrial companies or consulting agencies; and (iii) governmental or environmental organizations – not having the same requirements in terms of scientific validity (e.g. peer-reviewed), the tenth category indicates the development background and origins of the C-indicators.

## 5.2. STATISTICAL ANALYSIS OF EXISTING C-INDICATORS AT THE MICRO LEVEL

The overall distribution of the 55 sets C-indicators have been first analyzed in the literature section. A more refined analysis of their repartition within the aforementioned categories is now given in Table 5. Particularly, a focus is made here on the 20 sets of C-indicators available at the micro level of the CE to examine more in-depth their distribution across the proposed categories. The view provided by the synthesis and organisation of C-indicators through the present taxonomy gives indeed some interesting trends that deserve to be emphasized, for instance to identify some lacks among this cluster of C-indicators:

- Regarding the CE loops considered by reviewed micro-level C-indicators (category #2), the majority of them (90%) encompasses recycling loops, while 65% considerer remanufacturing activities and/or reuse loops, and less than half of them – 45% – take explicitly into consideration all the main CE loops (i.e. prolong/maintain, remanufacturing/reuse, and recycle) within the same and consistent indicators set. Even if these C-indicators at the micro-level do not include all the aspects of the CE, they tend to encapsulate more than the recycling option. By comparison, macro-level C-indicators, mainly developed in China, have a stronger focus on recycling than on other CE loops.

- In connection with the circularity performance (category #3): 80% of the C-indicators at the micro level of the CE evaluate an intrinsic circularity. 40 % examine directly the impacts on sustainability aspects induced by the circularity of tangible goods. Only 20% include both – i.e. inherent and consequential circularity – simultaneously within the same C-indicators framework. Note that when considering the circularity effects on sustainable development, most of the C-indicators depict economic and environmental impacts, social consequences remaining barely addressed. This missing dimension is an issue often highlighted within SDI framework, according to Singh et al. (2012): "Only few of them have an integral approach taking into account environmental, economic and social aspects. In most cases the focus is on one of the three aspects". As such, Geng et al. (2012) called for a more systematic evaluation system that integrates and harmonizes relationships between indicators of environmental, economic, and social development so that they could effectively supplement one another.

- In terms of the (retro- or pro-) perspective aspects of C-indicators (category #4), 8 sets of C-indicators out of 20 are dedicated to assess a potential circularity while 12 out of 20 are designed to deliver information on an effective – intrinsic or consequential – circularity. Note that one could make use of these 12 C-indicators sets to project on a hypothetical circularity levels. More interestingly, when crossing categories #3 and #4 it has been found that a very few number of micro-level C-indicators attempt to evaluate the potential impacts of CE loops on the sustainability performance i.e. by attempting to predict the economic or environmental benefits of circularity.

- Concerning the dimensionality, 60% propose a single indicator that aggregates the circularity performance at the micro scale, summarizing therefore several facets of the CE into a one-dimension information, which could be arguable (Cayzer et al. 2017). In fact, there is no existing standardized method to aggregate the performances of all the CE loops into a single indicator (Elia et al. 2017).

- Only 3 C-indicators sets among the 20 reviewed here at the micro level are designed for sector-specific purposes. Most of them – i.e. 17 out of 20 - are quite generic in the way they could be applied in a diverse range of products. Yet, these micro-level C-indicators are still in a pilot phase, and even if they can claim a certain transversality, most of them have been solely applied and tested on one specific product or industrial sector.

- Last but not least, an interesting fact is that almost half of these C-indicators – 45% – are linked to a computational tool, making their application and implementation more convenient for practitioners. By comparison, at the macro level of the CE implementation, the wide majority of C-indicators framework are still embodied in a textual format.

This systemic demarcation of C-indicators and their mapping through the developed taxonomy aims not only at highlighting current limitations but also at orienting future research to fill these gaps, as developed in sections 6 and 7.

Table 5 – Repartition of C-indicators into the main categories of the proposed taxonomy
(numbers in brackets indicate the number of C-indicators fitting a given criteria)

| Categories | Micro (out of 20) | Meso (out of 16) | Macro (out of 19) |
|---|---|---|---|
| Loops | recycling (18)<br>reuse/reman (13)<br>maintenance (9)<br>all (9) | recycling (16)<br>reuse/reman (12)<br>maintenance (7)<br>all (7) | recycling (18)<br>reuse/reman (10)<br>maintenance (6)<br>all (5) |
| Performance | intrinsic (16)<br>impact (8)<br>both (4) | intrinsic (9)<br>impact (11)<br>both (4) | intrinsic (17)<br>impact (15)<br>both (13) |
| Perspective | potential (8)<br>effective (12) | potential (9)<br>effective (8) | potential (2)<br>effective (17) |
| Dimensionality | single (12)<br>multiple (8) | single (5)<br>multiple (11) | single (1)<br>multiple (18) |
| Transversality | generic (17)<br>sector-specific (3) | generic (14)<br>sector-specific (2) | generic (18)<br>sector-specific (1) |
| Format | computational tool (9)<br>textual format (11) | computational tool (4)<br>textual format (12) | computational tool (0)<br>textual format (19) |

### 5.3. SELECTION TOOL: THE C-INDICATORS ADVISOR

In the literature related to eco-design tools, additionally to the developed taxonomies, authors have proposed diverse ways to identify the most relevant tools for a specific context, for instance, through multi-dimensional graphs (Bovea and Pérez-Belis, 2012), decision tree or associated online tool (Rousseaux et al. 2017). Here, the knowledge captured through this analysis and classification of C-indicators was synthesized in an Excel spreadsheet, used for developing a selection tool of C-indicators. The selection tool has been designed using Microsoft Excel software so that it can be disseminated and updated easily. "The C-Indicators Advisor" is indeed an Excel-based tool with macro enabled which is linked to the database of 55 sets of C-indicators classified according to the proposed taxonomy. Snapshots of this tool are given in Figure 4. The goal of this selection tool is to support the users in identifying and selecting the most appropriate circularity indicators in line with their requirements. It is mainly intended to industrial practitioners, decision-makers and policy-makers working in CE projects. But it remains accessible to everyone – novice or expert – interested in the circular economy implementation, e.g. in order to discover the possible contributions of C-indicators and how they can be used in practice.

In the input interface of the Excel file, eight questions are asked to direct the users towards the most suitable C-indicator(s) and related assessment framework, similarly to an expert system based on eight questions. Selection criteria are the following: i) level of measurement; ii) circularity perspective;

iii) circularity performance; iv) circularity loop; v) dimensionality; vi) usages and purposes; vii) transversality; viii) type and format. Once the query is completed, a click on the round logo at the top of the Excel spreadsheet, as illustrated in the snapshot of the Figure 4, will launch the search. Then, the tool directs the user automatically to the results table of recommended C-indicators. The advisor matches and selects the indicators to display according to an advanced filtering system – using Excel macros – that linked the query inputs to the organized databank of C-indicators.

In outputs, appropriate indicator(s) are identified and the following information is displayed: a) C-indicator name; b) working principle; c) details about the systemic level; d) details about the kinds of circularity; e) details about the dimensionality and unit; f) data required to compute the indicator; g) possible useful usages; h) authors and references; j) internet access link. Interestingly, a direct internet access link to each of the recommended C-indicators and their associated assessment framework (e.g. formulas to compute, web-based tool) is indicated, to get further details and, if relevant, to start experimenting and implementing such indicator(s).

Note that this selection tool of C-indicators is flexible in the way the databank is not frozen and may be easily updated. As such, it is possible to contribute in return to the tool development, enrichment, or consolidation, e.g. if researchers, industrialists or policy-makers are aware of, have tested, or are developing (new) C-indicators that are not inventoried yet in the actual databank. Indeed, a key challenge is to succeed in maintaining the databank up-to-date, regarding the increasing number of studies and articles published in relation to the CE. Last but not least, two complementary 2-minute videos have been recorded and put online to (i) explain simply how the selection tool works (here is the link of the tutorial video: https://youtu.be/nRNbWyHRzic) and to (ii) illustrate the use of this tool through an industrial example (here is the link of the case study showing the application of the tool for identifying appropriate C-indicators in an industrial context: https://youtu.be/kd51SsX0Be4).

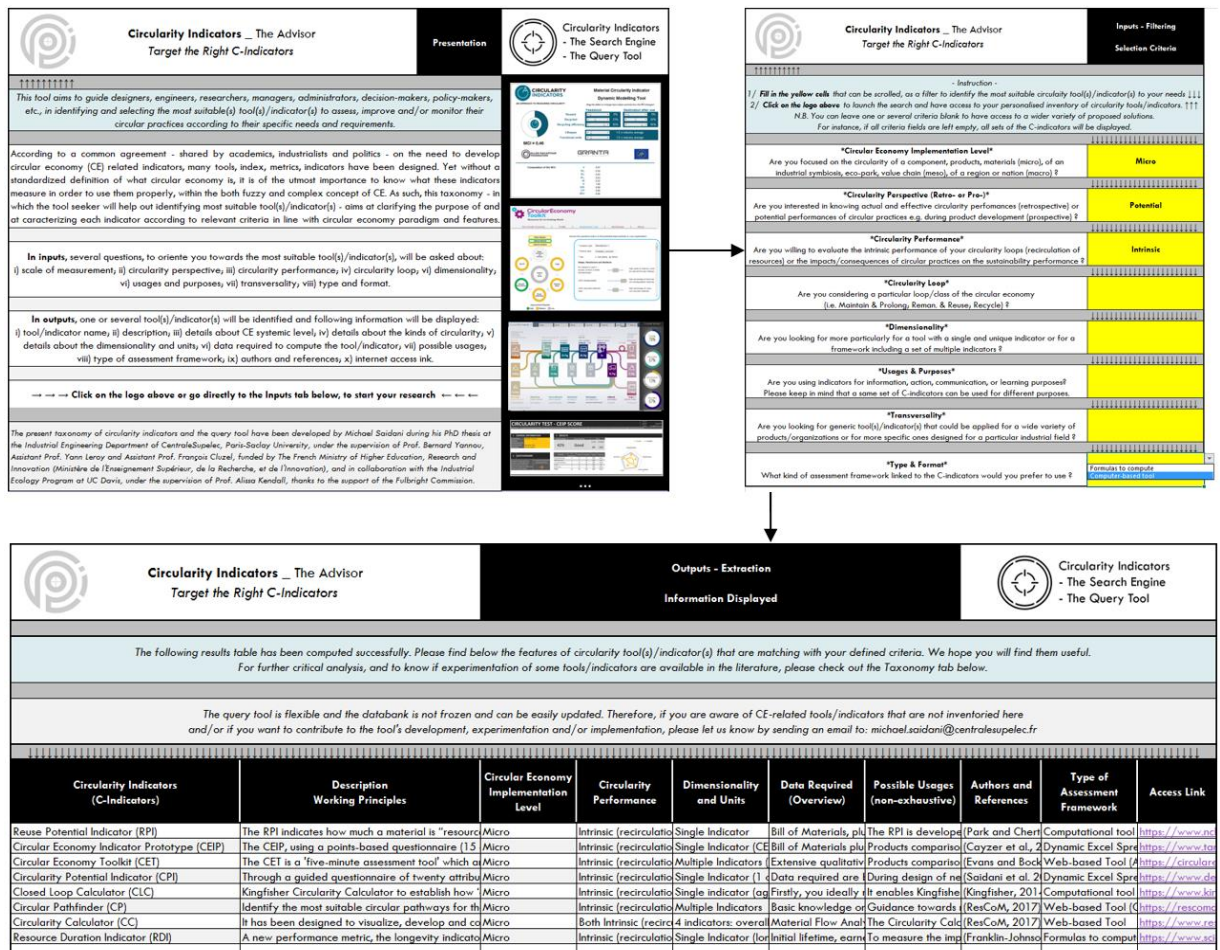

Figure 4 – Overview of the selection tool: "The C-Indicators Advisor"

Supplementary material and data exposed in this article, including the complete taxonomy of C-indicators and its associated Excel-based selection tool, can be found in the online version at [insert doi] (e.g. doi:https://doi.org/10.1016/j.jclepro.2017.11.064)

## 5.4. USE CASES AS A FIRST VALIDATION OF THE PROPOSED TAXONOMY

A first practical validation of the developed taxonomy is proposed by using its selection tool. The objective here is to check its robustness and contributions on the identification of appropriate C-indicators, based on use cases focused on the micro level of the CE – with the data published in literature – exploring how C-indicators can help (re-)designing better circular and sustainable products (e.g. used starter engines, prototype tidal energy device, or catalytic converter). More precisely, as we claim a need-based taxonomy, particularly driven by industrial needs at the CE micro-level, we are wondering whether: (i) the selection tool and associated taxonomy recommend the same C-indicators that are used in published case studies; and (ii) there is any other complementary set of C-indicators that would be also appropriate regarding to the initial purpose of a given case study.

Seven published use cases of C-indicators at the micro level of the CE – in which, one or several C-indicators are tested or used – have been identified to experience the proposed taxonomy and its associated selection tool, as illustrated in Table 6. The first column indicates the industry, product or material for which the circularity is measured. The second column specifies the objectives and purposes behind the use of C-indicators for each case study. The column three outlines the C-indicators originally considered and used in the case study. After translating the needs and requirements describing each case study into query inputs of the selection tool as indicated in the fourth column, C-indicators recommended are displayed in the fifth column.

Table 6 – Use cases of C-indicators at the micro level of the CE

| Case study and references | Overview of the initial objectives, needs and/or requirements | C-indicators used in the initial study | Query entered | C-indicators found by the advisor |
|---|---|---|---|---|
| Wine industry Acampora et al. 2017 | To measure circular practices, considering notably the recycling and reuse of secondary raw materials. | 3: **CEI**, **MCI**, **RPI** | Micro AND Recycle AND Reuse | 10: CC, **CEI**, CEIP, CET, CPI, IOBS, **MCI**, PCM, **RPI**, RDI |
| Mobile phone and precious metals Franklin-Johnson et al. 2016 | To enable managers to control the three longevity drivers: product use, refurbishment, recycling. To maximize resources exploitation through all the CE loops. | 1: **RDI** | Micro AND Potential AND All the loops | 6: CC, CEIP, CET, CLC, CPI, **RDI** |
| Plastic waste treatment Huysman et al. 2017 | To quantify the CE performance of different plastic waste treatment options, considering the environmental benefits. | 1: **CEPI** | Micro AND Actual AND Impact AND Recycle | 5: CEI, **CEPI**, CI, EVR, PCM |
| Used starter engines Linder et al. 2017 | To measure economic value capture through remanufacturing, reuse and recycling. | 1: **PCM** | Micro AND Actual AND Impact | 7: C2C, CEI, CEPI, CI, EVR, IOBS, **PCM** |
| Widgets EMF, 2015 | To compare the circularity of two products, considering products lifetime, and materials recycled or reuse. | 1: **MCI** | Micro AND Actual AND Intrinsic AND Generic | 7: C2C, CI, EOL-RRs, IOBS, **MCI**, RIs, RRs |
| Prototype tidal energy device Walker et al. 2018 | To compare the effectiveness of different material efficiency strategies and the correlation between product circularity and the environmental efficiency. | 3: CEIP, CET, MCI *in combination with LCA indicators* | Micro AND Potential AND Impact | 1: CC |
| Catalytic converter Saidani et al. 2017 | To evaluate circularity potential improvement during design and development process. | 3: **CEIP**, **CET**, **CPI** | Micro AND Potential AND Generic | 7: CC, **CEIP**, **CET**, CP, **CPI**, RDI, RPI |

For the mobile phone, plastic waste treatment, used starter engines and widgets case studies, the process is the following: one new C-indicator is developed and experimented on a specific use case that particularly fits with the indicator scope and purpose. On the other hand, the wine industry and prototype tidal energy device case studies seem more relevant here in the way the authors selected several C-indicators as relevant for their specific use cases among the sets of C-indicators they have initially identified and reviewed. For instance, in the wine industry case study, three C-indicators (MCI, CEI, RPI) have been selected as suitable out of the eight identified (CEPI, RPI, CEIP, CET, CET, MCI, EISCE, FCIM); and for the prototype tidal energy device case study, three were selected (MCI, CET, CEIP) among the six identified (MCI, CET, CEIP, CEI, C2C, VRE). The acronyms of all these C-indicators are listed in Appendix A

In most cases (6 out of 7), the C-indicators initially used have been also advised by the selection tool and supplementary indicators have has been suggested as well, which might have been insightful for these studies. On the other hand, the "prototype tidal energy device" case study (Walker et al. 2018) highlights the lack of multi-dimensional indicators considering both product circularity and sustainable performance within the same framework. Mathematically, regarding the combinatory aspects of the query tool, the approximately 300 possible pathways through criteria combination - among the 50+ sets of C-indicators inventoried - ensure a rapid convergence towards the most suitable C-indicator(s).

## 6. DISCUSSION

### 6.1. GAPS FILLED AND REMAINING LIMITATIONS IN THE MEASUREMENT OF THE CE

The identification and classification of available C-indicators allow to get a comprehensive and updated overview of the progress made on the circularity assessment, as well as to comment on the gaps filled in last few years (e.g. the measurement of CE at a micro level) and on the remaining challenges to orient future research (e.g the uptake of C-indicators by industrialists, or the issue of data availability to compute the indicators). On this basis, this tool seeker can serve the proper dissemination of appropriate C-indicators to monitor and support the CE transition in industry and policy making. Moreover, the potential complementarity or supplementarity between existing C-indicators is a point that would require further discussion and analysis. Also, the question of how indicators could complement one another has indeed still not been addressed satisfactorily.

#### 6.1.1. Progress at the micro-level and complementarity between C-indicators

Our study shows that previous statements advancing that few C-indicators are situated at the micro-level of the CE are somehow no longer true. For instance, in articles published in 2017, it has been said that "a deep research on CE assessment and indicators is still lacking, in particular on the micro level" and that "few studies are focusing on how to measure effectively the circularity level of a product, a supply chain or a service" (Elia et al. 2017), or that the evaluation of product circularity performance is a barely addressed topic (Saidani et al. 2017). Actually, in line with Walker et al. (2018) who mentionned a growing number of C-indicators at the micro level, our systematic review inventories 20 C-indicators at the micro level of the CE. Nonetheless, many of these C-indicators are under development and still in the pilot phase (Walker et al. 2018). According to Acampora et al. (2017), research about indicators for measuring the application level of CE strategies is still in its earliest phase, particularly on the micro level. This low degree of maturity (combined with a high degree of genericity) could be an explanation of their low degree of adoption in industrial practices (assumption based on the extrapolation of the scarce implementation of eco-design tools or sustainability indicators in industry, discussed in the scientific literature). Even if some progress has been and are currently done at this micro level, we believe the call made by Elia et al. (2017) "for further research about more effective CE strategies evaluation" remains relevant. More concretely, some existing and generic C-indicators at the micro level could serve as a suitable basis for the development of new ones more adapted for a specific context. For instance, Verberne (2016) developed a sector-specific indicators set for the building industry: the Building Circularity Indicators (BCI) based on modifications made on the Material Circularity Indicator (MCI) created by the Ellen MacArthur Foundation (EMF, 2015), facilitating as such its use for industrialists from the building sector, and demonstrating C-indicators can be built on one another.

Additionally, Elia et al. (2017) add no single existing indicator encompasses all the requirements of the CE paradigm. To them, "focusing on one single dimension of the CE (e.g. resource use) represents a limitation in the assessment of CE models, leaving other important factors, such as emissions and energy use". Only few of the C-indicators attempt to provide a more holistic approach taking into account both intrinsic circularity and the effects of this circularity e.g. on the three pillars of sustainable development. On this basis, coupled approaches mixing several C-indicators appear as a solution for an augmented measurement of the circularity performance. For instance, Figge et al. (2018) proposed a two-dimensional indicator, combining a longevity indicator – capturing how long product systems retain resource materials – with a circularity one – quantifying the number of times that a resource is passing through different phases in a value chain – in order to inform better decision making in the sustainable management of resource use. Pauliuk (2018) also emphasized that physical circularity indicators (e.g. the MCI, C2C or CEIP) can be complemented by monetary ones (e.g. the PCM, CEI, or EVR). The Ellen MacArthur Foundation (EMF, 2015) completed its MCI with environmental indicators such as water and energy consumption or greenhouse gas emissions to add a sustainable component when assessing the inherent circularity of materials. The comparison of C-indicators with LCA results may indeed reveals potential trade-offs e.g. between the goals of resources circularity and reducing environmental burden (Walker et al. 2018). Geissdoerfer et al. (2017) remind that in some cases, improving the intrinsic circularity performance might result in a negative environmental impact along the life cycle. Furthermore, the best end-of-life pathway may also vary when looking at the cost or at social component. That is why Figge et al. (2018) encourage further research on the combination between circularity measures and life cycle sustainability indicators. Finally, consequential LCA – contrary to the commonly used attributional LCA - is another possible solution still barely explored to evaluate the implementation of future CE projects. According to ScoreLCA (2015), "this method is capable of taking into account market evolutions to evaluate the environmental consequences of developing a new system or making a precise decision. By studying the environmental impacts associated with the implementation of a recycling loop or with the substitution of raw materials by recycled materials, it is possible to evaluate the effect this evolution might have on the environment or the market". Yet, the application of consequential LCA demands an important knowledge and numerous data related to the evaluated sector.

### 6.1.2. Current limits and potential solutions: data issue and industrial uptake

Wisse (2016) depicted an overview of knowledge gaps and shortcomings in the CE assessment literature, including a lack of: (i) knowledge and best practices of C-indicator frameworks; (ii) stakeholders' engagement in the design process of indicator frameworks; and (iii) CE indicators representing holistic fields. As similar challenges are found and have been extensively discussed in the fields of eco-design tools or sustainability indicators, CE researchers – ideally together with practitioners – should consider this existing literature in order to anticipate and overcome the identified barriers so as to facilitate the effective implementation of C-indicators in industrial practices. In fact, Rossi et al. (2016) explored the main barriers that prevent the implementation of eco-design approaches in industrial companies, and proposed possible strategies to overcome these barriers. In line with Bovea and Pérez-Belis (2012), most of the eco-design tools are not applied in a systematic way in companies due to their complexity, the time required to implement them and the lack of environmental knowledge. Park and Kremer (2017) remind that companies need to understand the relevance and potential benefits of environmental sustainability indicators to use them in the management of their operations. Yet, they state that "the lack of information with regards to the utility of indicators and the technical and theoretical orientation of indicators hamper their implementation in practice." Park and Kremer (2017) conducted thus an industrial survey on the utilization and utility of environmental sustainability indicators. As the research on C-indicators is still in development, a similar study, e.g. by using (an adapted version of) the framework they proposed, as exposed in Table 7, may be relevant to get a higher accuracy on the degree of awareness, interest and use of current C-indicators by industrialists.

Table 7 – Framework to evaluate the utilization and utility of indicators (Park and Kremer, 2017)

| Main criteria | Sub-criteria | Description | Input values |
|---|---|---|---|
| Utilization: current and future usage of an indicator | Used in practice | Current usage of an indicator | 1: not used; 2: in adoption phase; 3: currently used |

|  | Future implementation | Likelihood of implementing an indicator in the future | 1: no; 2: yes |
|---|---|---|---|
| Utility: inherent value and feasibility of an indicator | Usefulness | Perceived economic and operational value of an indicator | 1-5, with 5 being the most useful |
|  | Practicality | Perceived cost and time to learn and implement an indicator | 1-5, with 5 being the most practical |

Another key challenge to the proper computation of C-indicators is the need for various and important quantity of data all along the value chain. Much of this information is difficult to obtain and must be provided by the actors in the product chain itself (Potting et al. 2016). The data issue is indeed a major barrier to a wider use of indicators in companies due to the time and cost needed to collect them, the lack of information exchange in businesses, as well as confidential aspects (Birat, 2012). As such, special focus should be made on the data required to feed the indicators (Lützkendorf and Balouktsi, 2017). Furthermore, to Geissdoerfer et al. (2017), measurement as a means of improvement and optimization is still very much in an experimental phase, but it should increasingly be supported by the evolution of digital technologies, such as the Internet of Things. This could lead to the availability of completely new data sets, especially at the micro level of circularity, to assess the circularity performance of products, components and materials through the entire lifecycle. Currently, at the macro level, e.g. at the European level, a lot of relevant data for the circular economy are available thanks to the direct involvement of key data providers like Eurostat, the Joint Research Centre or the EEA (EC, 2015b).

## 6.2. AREAS OF IMPROVEMENT AND FLEXIBILITY OF THE PROPOSED TAXONOMY

The ten proposed categories and their associated criteria to classify existing C-indicators do not claim to be completely exhaustive, but rather to be a practical, usable and understandable way to find out an appropriate set of C-indicator for a given context. Indeed, the proposed categories encompass the main CE features (categories #1 to #4), the possible use of C-indicators (categories #5 and #6) and the key characteristics of their associated assessment framework (categories #7 to #10), allowing therefore a clear and rapid differentiation between C-indicators. Nevertheless, one could advance other possible – complementary or supplementary – categories to sort them out:

− The EEA (2016) suggested the measure and reporting of the degree of circularity achievements should be specified throughout the life cycle of products or systems, that is to say on the following stages: design (e.g. easy of disassembly), production (evolution of the overall (primary, secondary) use of materials), consumption (lifespan, use intensity), end-of-life (volume of landfill evolution).

− Additionally, at the micro level of CE implementation, to facilitate the integration of C-indicators in the industrial design and development process, it could be interesting to inform on which steps certain C-indicators can provide guidance and recommendations – e.g. on project scoping, concept definition, design definition, or product implementation as proposed by the ResCoM project (2017). CE-related tools and indicators available on the ResCoM platform have also an indication about their preparation, calculation and implementation time.

− ScoreLCA (2015) indicates a classification of material loops in three categories: (i) closed loops (short and mainly B2B); (ii) open loops (longer and mainly B2C); and (iii) cascade recycling like downcycling that considers the quality of recycled materials, which can therefore complete resource-oriented indicators mainly focused on the quantity of materials (Elia et al. 2017).

− In analogy to thermodynamics, it could also be relevant to indicate the extensive or intensive properties of C-indicators, notably at the meso and macro levels of CE implementation. While intensive indicators are independent of the size of the system, the value of extensive indicators depends of the system size. In order to make indicator results better comparable across countries, regions, cities or across different industrial sectors, intensive indicators are preferable (Eisfeldt, 2017) and extensive ones need to be normalized.

According to the original use of taxonomies in biology and natural sciences, Davidson (1952) reminded "the principles of taxonomy have not always been constant, they have changed as the objectives of taxonomy have altered through the years". At first, their major objectives were to enable the identification and classification of species. Then, it was to determine the interrelationships between identified species. As such, and by analogy with this, we can argue the future steps will be to establish further links and correlations between existing C-indicators. Eventually, one has to bear in mind such characterisation of C-indicators has to be questioned and updated on a regular basis because of the complex and rapid dynamics governing the CE transition (EEA, 2016). According to the EEA (2016), a CE monitoring framework should be flexible to maintain the indicators effectiveness throughout the evolution of the transition. Indeed, any indicator set – particularly in the fields of sustainability and circularity – should be adaptive enough to reflect the varying and time-evolving stakeholders' needs (Lützkendorf and Balouktsi, 2017). Against this background, and in line with the discussion in Section 6.1, a next update of the proposed taxonomy could be to add a compatibility matrix between the C-indicators e.g. based on their associations and/or occurrences in published use cases. Such information would enrich the taxonomy by offering an augmented orientation in the selection of an appropriate set of C-indicators.

## 7. CONCLUSION AND PERSPECTIVES

One of the core questions around the CE is how to measure its progress and performance at different levels, regarding how complex and fuzzy this CE concept can be. As a response to the need of monitoring the CE transition, an increasing number of attempts to develop circularity indicators have been noticed in the last few years, covering more or less the multi-facets of the CE. In this article, the taxonomically sound characterisation and classification of 55 sets of C-indicators brings some clarity on their purposes and therefore support their appropriate use and dissemination, notably thanks to a user-friendly selection tool associated to the database of these C-indicators. Through the developed taxonomy, the organised categorisation of C-indicators can assist industrial practitioners and policy-makers who need to be informed to make decisions on CE-related projects. Indeed, without C-indicators it is difficult to draw any conclusions, and having the wrong C-indicator could lead to non-appropriate conclusion.

Limitations of the proposed taxonomy, as well as some improvement areas that need be investigated further have already been partly mentioned in the discussion section. Yet, further emphasis is placed here to expand and open up the discussion on three key perspective: (i) the advanced robustness of – existing and future – C-indicators; (ii) their enhanced adoption by industrialists to conduct CE strategies; and (iii) their contribution to catalyze the transition towards a more CE. As such, this article provides a baseline for new and upcoming investigations into the potential development and implementation of *ad hoc* C-indicators. The following sub-sections aim to guide more precisely future research on the measurement of the CE performance.

### 7.1. FURTHER EVALUATION OF EXISTING C-INDICATORS

Future work should evaluate and judge more objectively the definition, relevance and scientific soundness of C-indicators, so that one can have more trust and confidence in their use. Delivering insights at the question of which criteria to use to do so is an essential first step. According to the EEA (2003), a good indicator should: communicate in a sound way a simplified reality; match the interest of the target audience; be attractive to the eye and accessible; be easy to interpret; be representative of the issue or area being considered; invite action: show developments over a relevant time interval; go with a reference value for comparing changes over time; be comparable with other indicators that describe similar areas, sectors or activities; and be scientifically well-founded. Weiland (2006) proposed methical requirements for sustainability indicators such as: having a clear rationale; representing an adequate image of complex system; having face validity; being specified clearly; being repeatable. Moreover, identifying the sources and levels of uncertainties (e.g. coming from data quality, assessment methodology) for such indicators are of paramount importance.

To choose indicators related to resource efficiency, the European Commission (Eisenmenger et al. 2016; EC, 2009) used the following criteria: policy relevance; coverage of all relevant categories and resources; coherence and completeness; transparency of trade-offs and negative side effects such as burden shifting; applicability to different levels of economic activities. Other lists of criteria for selecting indicators have been put forward, notably by managers or consulting companies. For instance, the consulting agency Deloitte (BioIS, 2012) has recommended the usage of RACER criteria (relevant, acceptable, credible, easy, robust) to evaluate indicators' suitability. Other efficient mnemonics ways are usually used in companies to define and select indicators, inspired from managerial best practices such as SMART (specific, measurable, achievable, relevant, timed) or CREAM (clear, relevant, economic, adequate, monitorable). These acronyms represent commonly used criteria for performance indicators. They are widely used in the manifold sectors to provide 'rule of thumb' guidance to managers identifying most suitable indicators. Importantly, it is widely acknowledged that indicators are only relevant and useful if they fit the user's needs (Bouni, 1998). More recently, some authors provide more particular guidance and recommendations for the development of C-indicators (Iacovidou et al. 2017; Saidani et al. 2017a). Using such criteria and framework can therefore be meaningful during the definition, development and setup of future C-indicators, as well as in the validation of newly proposed C-indicator sets.

### 7.2. FURTHER UPTAKE OF C-INDICATORS BY INDUSTRIAL PRACTITIONERS AND POLICY MAKERS

By shedding a light on a wide variety of exisiting C-indicators in an organized and understandable manner, we argue this study can contribute in their appropriate use in practice. Indeed, the proposed taxonomy can be a first step in making practitioners aware of the opportunities offered by the application of suitable C-indicators and therefore could support their effective uptake by industry and use by policy makers in the setup and monitoring of CE-related regulatory frameworks (Tecchio et al. 2017). As the CE transition process consists of means (e.g. product chain partners, knowledge development), activities (e.g. knowledge exchange, experimentation of new business models) and achievements (e.g circularity of resources, lowering environmental impact) (EEA, 2016; Potting et al. 2016), information given by C-indictors can serve as a useful binder e.g. for managers in charge of monitoring the transition towards more CE practices. Indeed, in the transition movement to the CE, indicators are needed to track progress and to provide direction on where to go next. Interestingly, the further development of sector-specific C-indicators can concretely foster their adoption, e.g. in the building sector (Núñez-Cacho et al. 2018; Verberne, 2016).

In this line, to make this circular vision more straightforward and shared by decision-makers, including policy-makers as well as industrial practitioners, efforts must be done on: the appropriate level of intelligibility of C-indicators (e.g the indicators discretization) in accordance to their main recipients; the simple translation of the information given by a C-indicator into precise actions or practical recommendations; the correlation between circularity heuristics and more tangible impacts; the integration of C-indicators e.g. in the industrial development process to design more circular products. Also, communication on best practices or successful examples of how C-indicators have helped managerial activities to orientate actions in CE projects, as well as new experimentations of C-indicators for steering circular strategies, should be foster to lead and inspire this shift towards a more CE. Finally, as mentioned in the previous section, making C-indicators more transparent and trustworthy e.g. in anticipating the environmental or economic performance and thus enlightening decision-making (Thomas and Birat, 2013), will make them certainly more applicable in return.

### 7.3. FURTHER IMPLEMENTATION OF THE CE

To put things in perspective, one has to bear in mind C-indicators are solely one element in the overall process of the CE transition. In fact, even if this work offers a valuable framework for future research related to the measurement, improvement and monitoring of the CE performance, it is important to remind, in line with the EMF (2013), that the successful implementation of CE models relies on the synergy between key building blocks including product design, new business models, reverse logistics, enablers and systems conditions. From that standpoint, C-indicators can be considered as interesting enablers of the move to a more CE. Yet, the information provided by those C-indicators has to be translated into suitable actions for managing the CE transition. As such, other methods, tools and resources can complementary help the implementation of CE. For instance, published recently, the BS 8001:2017 is the first standard to guide organizations in implementing the principles of the CE. Globally, the implementation of CE strategies requires new organizational and business models, enhanced technologies (Hass et al. 2015), augmented know-how and shared knowledge (Park and Chertow, 2014), as well as a redefinition of industrial process and product innovations (EEA, 2016). And all these changes have to be economically, socially and environmentally sustainable to guarantee a successful implementation of the CE – effective and efficient – in the long run.

## REFERENCES (120)

# APPENDIX A - NOMENCLATURE OF THE C-INDICATORS REVIEWED

Table A.1 – List, acronyms and sources of the 55 C-indicators reviewed in the proposed taxonomy

| Acronyms | C-Indicators | Sources (authors and year) |
|---|---|---|
| ACT | Assessing Circular Trade-offs (ACT) | Circle Economy and PGGM, 2014 |
| BCI | Building Circularity Indicators (BCI) | Verberne, 2016 |
| C2C | Material Reutilization Part (C2C) | C2C, 2014 |
| CA | Circle Assessment (CA) | Circle Economy and PGGM, 2014 |
| CAT | Circularity Assessment Tool (CAT) | PGGM, 2015 |
| CBT | Circular Benefits Tool (CBT) | Advancing Sustainability LLP, 2013 |
| CC | Circularity Calculator (CC) | ResCoM, 2017 |
| CECAC | Circular Economy Company Assessment Criteria (CECAC) | VBDO, 2015 |
| CEI | Circular Economy Index (CEI) | Di Maio and Rem, 2015 |
| CEII | Circular Economy Indicators for India (CEII) | Talwar, 2017 |
| CEIP | Circular Economy Indicator Prototype (CEIP) | Cayzer et al. 2017 |
| CEMF | Circular Economy Monitoring Framework (CEMF) | European Commission, 2017 |
| CEPI | Circular Economy Performance Indicator (CEPI) | Huysman et al. 2017 |
| CET | Circular Economy Toolkit (CET) | Evans and Bocken, 2013 |
| CETUS | Circular Economy Toolbox US (CETUS) | US Chamber Foundation, 2017 |
| CEV | Circular Economic Value (CEV) | Fogarassy et al. 2017 |
| CI | Circularity Index (CI) | Cullen, 2017 |
| CIPEU | Circular Impacts Project EU (CIPEU) | European Commission, 2016 |
| CIRC | Circularity Material Cycles (CIRC) | Pauliuk et al. 2017 |
| CLC | Closed Loop Calculator (CLC) | Kingfisher, 2014 |
| CP | Circular Pathfinder (CP) | ResCoM, 2017 |
| CPI | Circularity Potential Indicator (CPI) | Saidani et al. 2017 |
| DEA | Super-efficiency Data Envelopment Analysis Model (DEA) | Wu et al. 2014 |
| ECEDC | Evaluation of CE Development in Cities (ECEDC) | Li et al. 2010 |
| EISCE | Evaluation Indicator System of Circular Economy (EISCE) | Zhou et al. 2013 |
| EMCEE | Indicators for Material input for CE in Europe (IMCEE) | EEA, 2016 |
| EoL-RRs | End-of-Life Recycling Rates (EoL-RRs) | Graedel et al. 2011 |
| EPICE | Environmental Protection Indicators (EPICE) in a context of CE | Su et al. 2013 |
| ERCE | Evaluation of Regional Circular Economy (ERCE) | Chun-Rong and Jun, 2011 |
| EVR | Eco-efficient Value Ratio (EVR) | Scheepens et al. 2016 |
| EWMFA | Economy-Wide Material Flow Analysis (EWMFA) | Haas et al. 2015 |
| FCIM | Five Category Index Method (FCIM) | Li and Su, 2012 |
| HLCAM | Hybrid LCA Model (HLCAM) | Genovese et al. 2017 |
| ICCEE | Indicators for Consumption for CE in Europe (ICCEE) | EEA, 2016 |
| ICT | Circularity Indicator Project (ICT) | Viktoria Swedish ICT, 2015 |
| IECEE | Indicators for Eco-design for CE in Europe (IECEE) | EEA, 2016 |
| IECF | Indicators of Economic Circularity in France (IECF) | Magnier, 2017 |
| IEDCE | Integrative Evaluation on the Development of CE (IEDCE) | Qing et al. 2011 |
| IOBS | Input-Output Balance Sheet (IOBS) | Marco Capellini, 2017 |
| IPCEE | Indicators for Production for CE in Europe (IPCEE) | EEA, 2016 |
| IPCEIS | Industrial Park Circular Economy Indicator System (IPCEIS) | Geng et al. 2012 |
| MCI | Material Circularity Indicator (MCI) | EMF, 2015 |
| MRCCEI | Measuring Regional CE–Eco-Innovation (MRCEEI) | Smol et al. 2017 |
| NCEIS | National Circular Economy Indicator System (NCEIS) | Geng et al. 2012 |
| PCM | Product-Level Circularity Metric (PCM) | Linder et al. 2017 |
| RCEDI | Regional Circular Economy Development Index (RCEDI) | Guo-Gand and Jing, 2011 |
| RDI | Resource Duration Indicator (RDI) | Franklin-Johnson et al. 2014 |
| RES | EU Resource Efficiency Scoreboard (RES) | Eurostat, 2015 |
| RIs | Recycling Indices (RIs) for the CE | Van Schaik and Reuter, 2016 |
| RP | Resource Productivity (RP) | Wen and Meng, 2015 |
| RPI | Reuse Potential Indicator (RPI) | Park and Chertow, 2014 |
| RRs | Recycling Rates (RRs) | Haupt et al. 2016 |
| SCI | Sustainable Circular Index (SCI) | Azevedo et al. 2017 |
| VRE | Value-based Resource Efficiency (VRE) | Di Maio et al. 2017 |
| ZWI | Zero Waste index (ZWI) | Zaman and Lehmann, 2013 |

Supplementary material and data related to this article, including the complete taxonomy of C-indicators and its associated Excel-based selection tool, can be found in the online version at [insert doi] (e.g. doi:https://doi.org/10.1016/j.jclepro.2017.11.064)